\documentclass[%
 aip,
 amsmath,
 amssymb,
 superscriptaddress,
 reprint,%
]{revtex4-2}
\usepackage{graphicx}
\usepackage{dcolumn}
\usepackage{bm}

\usepackage{upgreek}

\usepackage{xcolor}

\usepackage{hyperref}

\makeatletter
\newcommand{\manuallabel}[2]{\def\@currentlabel{#2}\label{#1}}
\makeatother

\begin{document}
\manuallabel{fig_saeulen}{S3}
\manuallabel{fig_abs}{S2}
\manuallabel{fig_qwp_charac}{S1}

\preprint{AIP/123-QED}

\title[The impact of metallic contacts on spin-polarized photocurrents in topological insulator $\text{Bi}_2\text{Se}_3$ nanowires]{The impact of metallic contacts on spin-polarized photocurrents in topological insulator $\text{Bi}_2\text{Se}_3$ nanowires}

\author{N. Meyer}
\email{nina.meyer@uni-greifswald.de}
\thanks{corresponding author}
\affiliation{Institute of Physics, University of Greifswald, Felix-Hausdorff-Str. 6, 17489 Greifswald, Germany}
\author{K. Geishendorf}
\affiliation{IFW Dresden, Institute for Metallic Materials, Helmholtzstra{\ss}e 20, 01069 Dresden, Germany}
\author{J. Walowski}
\affiliation{Institute of Physics, University of Greifswald, Felix-Hausdorff-Str. 6, 17489 Greifswald, Germany}
\author{A. Thomas}
\affiliation{IFW Dresden, Institute for Metallic Materials, Helmholtzstra{\ss}e 20, 01069 Dresden, Germany}
\affiliation{Institute of solid state and materials physics, Technische Universit{\"a}t Dresden, 01062 Dresden, German}
\author{M. M{\"u}nzenberg}
\affiliation{Institute of Physics, University of Greifswald, Felix-Hausdorff-Str. 6, 17489 Greifswald, Germany}


\begin{abstract}
Recently, a quantum phase, the topological insulator, has been vividly investigated in a variety of materials. Its unique bandstructure allows for optical generation and control of spin-polarized currents based on the circular photogalvanic effect. In this paper, we generate and distinguish the different photocurrent contributions via the polarization of the driving light wave. We discuss the helicity-dependent spin-polarized current and the polarization-independent thermoelectric current as spatially resolved maps, focusing on the influence of the topological insulator/metallic contact interface. We observe for both current contributions a significant enhancement of the current values at the topological insulator/metallic contact interface. In the case of the thermoelectric current the enhancement is localized at the center of the interface. The spin-polarized current reaches two extrema per contact, which differ by their sign and are localized nearby the contact edges.
We discuss the general behavior of the thermovoltage as a three-material Seebeck effect and explain the enhanced values by the acceleration of the photoelectrons generated in the space charge region of the topological insulator/metallic contact interface. Furthermore, we interpret the temperature gradient together with the spin Nernst effect as a possible origin for the enhancement and spatial distribution of the spin-polarized current.
\end{abstract}

\maketitle
\begin{figure}
\includegraphics[scale=1.18]{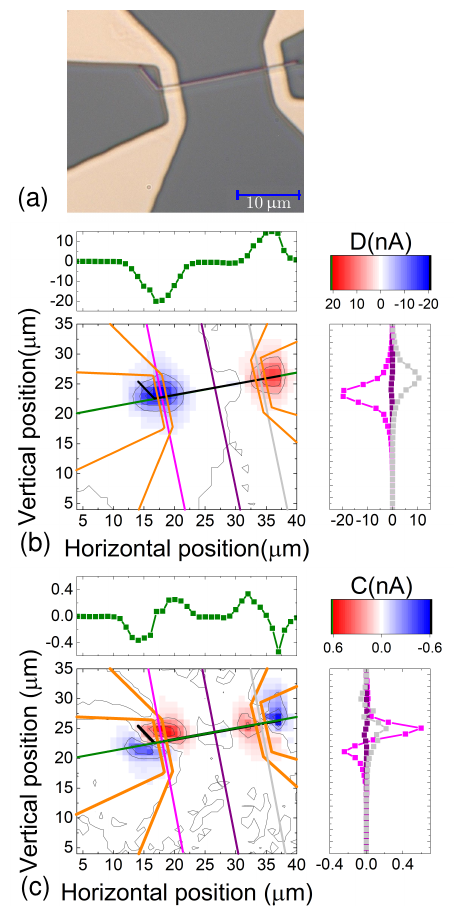}
\caption{\label{fig_14um} Photocurrent measurement on a $\text{Bi}_2\text{Se}_3$ nanowire with a $14\,\mathrm{\upmu m}$ gap with the drain (source) electrode on the right (left). (a) shows the micrograph of the measured area. The spatially resolved results for (b) the thermoelectric current D and (c) the spin-polarized current C  are presented, including contour plots along the nanowire (green lines), the contacts (pink and gray lines) and through the center of the nanowire (violet lines).}
\end{figure}
\begin{figure*}
\includegraphics[scale=1.18]{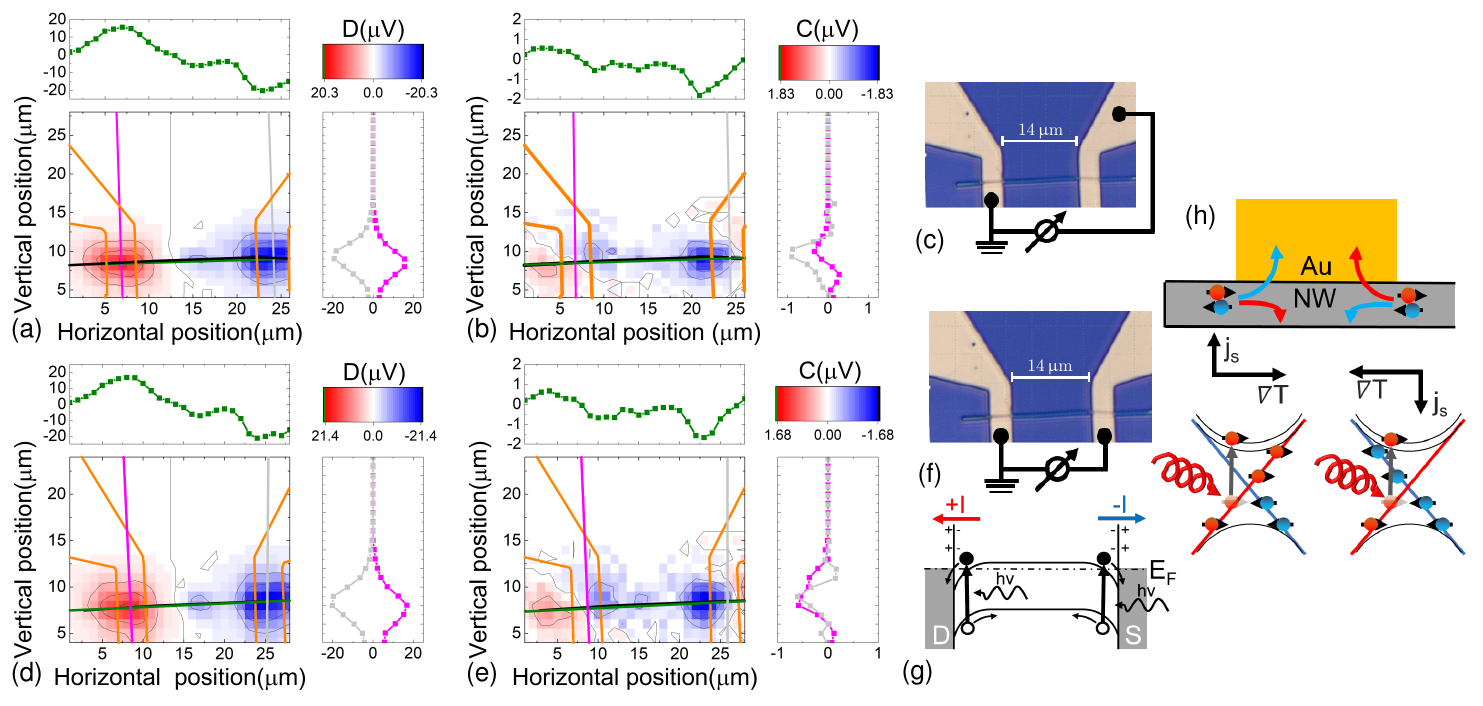}
\caption{\label{fig_14um_geom} Photovoltage measurements on a $\text{Bi}_2\text{Se}_3$ nanowire for different contacts. In (a), the thermovoltage D and in (b), the voltage C generated by the spin-polarized current for the axisymmetric contacts are displayed. In (c), the axisymetric bonding is marked in black on a micrograph of the sample. In (d), the thermovoltage D and in (e), the voltage C generated by the spin-polarized current are displayed for point-symmetric bonding. The bond position is marked in (f) on a micrograph of the sample. The contour plots along the nanowire (green lines) and along the contacts (pink and grey lines) are displayed next to the maps. In (g), the influence of the band bending at the contact/nanowire interface on the photoelectrons is illustrated. In (h), the two-step process for the spin accumulation at one contact is shown. The deflection of the spin-polarized electrons is sketched in the upper panel and the asymmetric population at the left and right contact edge is shown in the lower panel.} 
\end{figure*}

Recently, topological order gained a lot of attention among physicists after the discovery of topological insulators (TIs) \cite{Thouless1982,KaneMele2005,Koenig2007} in solid state materials. This additional state of quantum matter differs from trivial insulators by hosting a bulk energy gap, while the surface possesses gapless electronic states. The nanowires investigated in this paper consist of $\text{Bi}_{2}\text{Se}_{3}$. This material is a three-dimensional topological insulator with time-reversal symmetry\cite{Zhang2009, Xia2009}. Thus, the surface states of $\text{Bi}_{2}\text{Se}_{3}$ are helical and the spin degeneracy is lifted at the surface. Also the backscattering of the surface electrons is suppressed \cite{HasanKane2010,Hsieh2008}. Those three properties make $\text{Bi}_{2}\text{Se}_{3}$ a promising candidate for spintronics- or optoelectronics applications.
\newline The surface states make those materials suitable for polarization-sensitive detectors based on the polarization-dependent photovoltaic or photogalvanic effects \cite{Yan2014,Sharma2016}. Both effects are based on creating an asymmetric population of the spin-polarized surface states. Since the momentum and spin orientation are locked at the surface, the asymmetry in the surface state population generates a spin-polarized electrical current. When the asymmetry is generated by exciting the surface states with circular-polarized light, the effect is known as the circular photogalvanic effect (CPGE) \cite{Ganichev2003}. The CPGE  has been observed in several optoelectronic experiments, which demonstrate the direction control of spin-polarized currents by circular-polarized light in TI materials \cite{McIver2011, Kastl2015, Seifert2017}. The dimensions of those investigated TIs are in the  micrometer range. Little is known about those effects when the dimensions of TIs decrease towards the nanometer scale e.g. in nanowires. There is one example for all-optical control in $\text{Bi}_2\text{Te}_3\text{Se}$ nanowires by Seifert et al.\cite{Seifert2017}. They observe an enhancement of the THz amplitude in the vicinity of the gold contacts which they explain by a locally enhanced spin-polarized current. In a previous work we measured the polarization-dependent photocurrent while scanning the $\text{Bi}_{2}\text{Se}_{3}$ nanowire \cite{Meyer2020}. We observed that the polarization-independent current contribution is dominated by the Seebeck effect. We also detect a helicity-dependent current contribution caused by the CPGE. In both cases, we observed that the values of the contributions increase in the vicinity of the gold contacts, but we did not discuss the origin in detail. In this paper, we explicitly discuss the influence of the metallic contacts on the thermoelectric and spin-polarized current. We generate the two-dimensional scans for both contributions to discuss the origin of their enhancement close to the topological insulator/metallic contact interface.\newline
%
\begin{figure}
\includegraphics[scale=0.79]{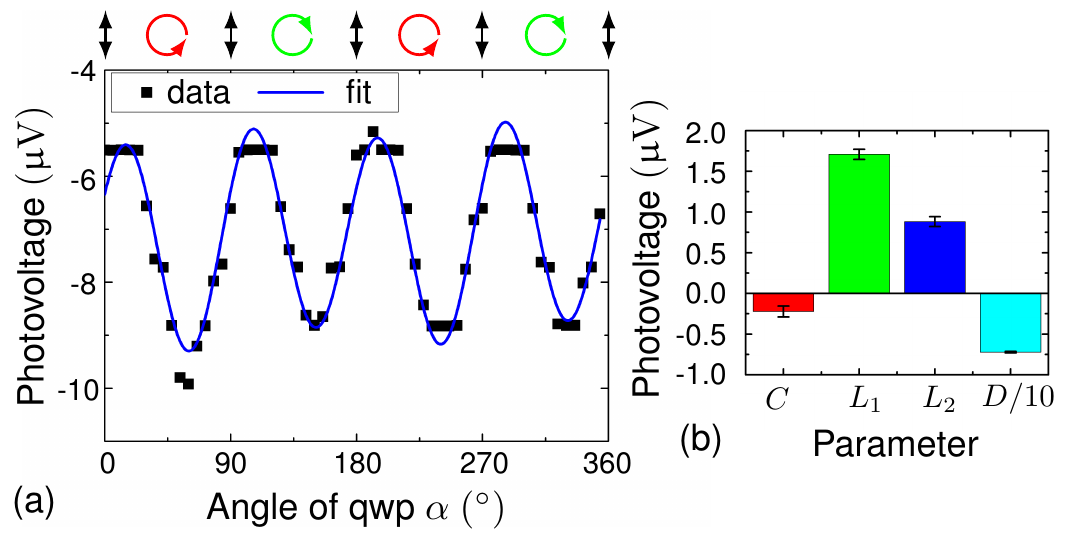}
\caption{\label{fig_example} Measured photovoltage $v\left(\alpha\right)$ for $\alpha=0,\Delta\alpha,\ldots,360$ with $\Delta\alpha=6^\circ$ at position (17,8) in FIG.~\ref{fig_14um_geom}(d). In (a), the measured data (black squares) and the fit function (blue line) are displayed. The arrows represent the polarization state of the laser light. In (b), the fitting parameters $C,L_1,L_2\text{ and } D/10$ and their uncertainties are displayed.}
\end{figure}
The photoelectric measurements in this paper are performed on $\text{Bi}_2\text{Se}_3$ nanowires synthesized by the gold-catalyzed vapor-liquid-solid method (for more details on the nanowire growth see Shin et al. \cite{Shin2016}). The nanowires are grown in [110] direction as single-crystal structures assuring a smooth surface. The nanowire width is in the order of $50\,\mathrm{nm}$ and the thickness spans from $50\,\mathrm{nm}$ to $150\,\mathrm{nm}$, thus hybridization of the topological surfaces states  can be excluded \cite{Gooth2014, Shin2016}. The nanowires are transferred mechanically to a Si(111) substrate with $100\,\mathrm{\upmu m}$ of $\text{SiO}_x$. The contact pads of $5\,\mathrm{nm}$ Cr underneath $50\,\mathrm{nm}$ Au are patterned on top of the nanowire with a spacing of $14\,\mathrm{\upmu m}$. 
The photoelectric measurements are recorded using a well established lock-in technique. The light source is a diode laser with a wavelength of $785\,\mathrm{nm}$ ($1.55\,\mathrm{eV}$) modulated at $77\,\mathrm{Hz}$ by a square-function generator. The laser beam is guided through a linear polarizer combined with a rotatable quarter-wave plate (qwp) for polarization control before impinging at an angle of incidence of $\theta=45^\circ$ on the sample surface. The laser light is focused down to a spot size of $\left(2.9\pm0.08\right)\,\mathrm{\upmu m}\times \left(3.4\pm0.12\right)\,\mathrm{\upmu m}$ on the sample surface. The measured photocurrent $j\left(\alpha\right)$ (photovoltage $v\left(\alpha\right)$) is mapped to the polarization state via the qwp rotational angle $\alpha$. Thus, the photocurrent is measured at fixed positions (k,j) for linear, left-circular and right-circular polarized light, scanning the sample surface vertically and horizontally by moving the laser across the sample surface with stepper motors. At the same time, the light intensity  reflected from the sample surface $I(\alpha)$ is recorded by a second lock-in amplifier via a photodiode. The qwp rotates at every laser spot position (k,j) from $\alpha=0,\Delta\alpha,\ldots,360^\circ$ with a step size of $\Delta\alpha=6^\circ$. Afterwards, the laser spot is moved in the vertical direction by $\Delta j=1\,\mathrm{\upmu m}$ to the next position  $(k,j-\Delta j)$. There, the photocurrent $j\left(\alpha\right)$ and the reflected intensity $I\left(\alpha\right)$ are measured again for a full qwp rotation. This generates a set of 60 current (voltage) values at each position (j,k) as presented in FIG.~\ref{fig_example}(a). \newline The data is evaluated by identifying the contributions in the measured photocurrent (voltage) by their polarization dependence. We use the ansatz suggested by McIver et al. \cite{McIver2011} that includes four contributions: 
\begin{equation}
j\left(\alpha\right)=C\sin(2\alpha)+L_1 \sin (4\alpha)+L_2\cos(4\alpha)+D\label{eq_current}.
\end{equation}
In our measurements, the laser light is linear polarized at $\alpha=0, 90, 180, 270 \text{ and } 360^\circ$, left-circular polarized $\sigma^+$ at $\alpha=45 \text{ and } 225^\circ$ and right-circular polarized $\sigma^-$ at $\alpha=135 \text{ and } 345^\circ$. The term $C\sin(2\alpha)$ in Eq.\eqref{eq_current} represents the spin-polarized helicity-dependent current. The amplitude $2C$ is the photocurrent difference between $\sigma^+$- and $\sigma^-$-light. The qwp phase is set so that $\sin(2\alpha)$ is zero for linear polarized light. The physical origin of this term is still under discussion. Measurements by Shalygin et al. on (110)-grown GaAs/AlGaAs quantum wells prove that the circular photon drag effect can cause a helicity-dependent current \cite{Shalygin2007}, while later measurements on exfoliated $\text{Bi}_2\text{Se}_3$ Hall bar devices by McIver et al.\cite{McIver2011} relate the helicity-dependent currents to the CPGE. The lifted spin degeneracy of surface electrons in a TI enables to generate an asymmetrical population of the spin-polarized surface states in the k-space by circular polarized light, since the optical selection rules for interband transitions have to be fulfilled. The asymmetric populated surface states cause an electrical spin-polarized current, since the spin and its momentum are locked on the surface of a TI. That exciting the spin-polarized surface states optically can lead to spin-polarized photocurrents has among others been demonstrated by Takeno et al. \cite{Takeno2018} by performing time-domain terahertz wave measurements and time-resolved magneto-optical Kerr rotation measurements. Hence, it is possible to generate a measurable helicity-dependent photocurrent due to the CPGE, which is spin-polarized \cite{Ganichev2003, Kastl2015, Okada2016, Plank2016, Braun2016, Seifert2017}. Therefore, we relate the first term $C\sin(2\alpha)$ in Eq.~\eqref{eq_current} to the CPGE following the notation established in the recent works \cite{McIver2011, Luo2017} and discuss the amount of spin-polarized current by displaying the amplitude $C$. \newline The second term $L_1 \sin (4\alpha)$ and third term $L_2\cos(4\alpha)$ in Eq.~\eqref{eq_current}, have a different frequency than the first term. Thus, they do not affect the first term. The third term $L_2\cos(4\alpha)$ represents the polarization-dependent absorption. The physical effects related to the two contributions are still under discussion both being sometimes related to the linear photogalvanic effect or the photon drag effect \cite{McIver2011, Olbrich2014, Plank2016B}. The last term $D$ is polarization-independent and can originate from the polarization-independent photogalvanic effect, the photon drag effect and, due to the laser light heating, the Seebeck effect. We observed in a previous work ref.~\cite{Meyer2020}, that the contributions of the polarization-independent photogalvanic and photon drag effect to the over all behavior of D seem to be small compared to the contribution of the Seebeck effect.\newline For data analysis, the function $j(\alpha)$ ($v(\alpha)$) is fitted to the experimental data to extract the parameters $C$, $L_1$, $L_2$ and $D$ at each position $(j,k)$ as in FIG.~\ref{fig_example}(a). The values for the four parameters and their uncertainties are displayed in FIG.~\ref{fig_example}(b). The parameters $C$ and $D$ are displayed in two dimensional maps as a function of the laser spot position, see FIG.~\ref{fig_14um}(b) and (c). The contact pad positions (edges are marked by orange lines) and the nanowire (black line) are determined from the reflectivity data and are in good agreement with the microscope images of the area of interest (see FIG.~\ref{fig_14um}(a) and FIG.~\ref{fig_14um_geom}(c) and (f)). In the following we concentrate on the parameters $C$ and $D$ to investigate the influence of the thermoelectric current on the spin-polarized photocurrent and their enhancement in the vicinity of the contact pads.\newline The two-dimensional map of the parameter D in FIG.~\ref{fig_14um}(b) shows that the thermocurrent reaches the largest values at the crossing of the nanowire and the contacts. This is visible in the contour plots in gray and pink in FIG.~\ref{fig_14um}(b). The contour plot (green line in FIG.~\ref{fig_14um}(b)) along the nanowire demonstrates that the increase is located at the metal contacts. Between the contacts the thermocurrent changes it sign and the slope of the thermocurrent is small, just like in ref.~\cite{Meyer2020}. When the laser partly illuminates the crossing of the nanowire and metal contact, the thermocurrent increases drastically until it reaches its largest value at a horizontal position of $35\,\mathrm{\upmu m}$. The origin of this enhancement is discussed later in the manuscript. 
\newline The current map of the spin-polarized current, represented by the parameter C, is displayed in FIG.~\ref{fig_14um}(c). C is constant in the middle of the nanowire where the contact pads can be neglected and thus the transport is entirely determined by the surface states. This matches prior research~\cite{Meyer2020}. When the laser spot illuminates the nanowire close to the contact pads, the spin-polarized current increases and reaches its extremal values nearby the contact edges. \newline The measurement was repeated on a second nanowire, maintaining the contact distance and laser conditions, but changing the contact geometry. The results for axisymmetric contacts or point-symmetric contacts are shown in FIG.~\ref{fig_14um_geom}. For the thermoelectric contribution, we observe for both contact geometries enhanced values at the crossing of the contacts and the nanowire. The thermovoltage in FIG.~\ref{fig_14um_geom}(a) reaches $D=15.13\,\mathrm{\upmu V}$ at the drain electrode and goes down to $D=0.85\,\mathrm{\upmu V}$ at position (12,7) on the nanowire, which qualitatively matches the findings for the thermoelectric current in FIG.~\ref{fig_14um}. In FIG.~\ref{fig_14um_geom}(b) and (e), the voltage C generated by the spin-polarized current for the two contact geometries are displayed. The voltage C reaches a minimum of $C=-1.84\,\mathrm{\upmu V}$ at (22,8) close to the source electrode and decreases to $C=-0.52\,\mathrm{\upmu V}$ at (14,8) at the nanowire center. Nearby the crossing of the contact pad and the nanowire, the voltage C reaches its extrema, which is in good agreement with the results observed in FIG.~\ref{fig_14um}(c). Switching between the axisymmteric and the point-symmetric contacts does not affect the shape of the distribution or the sign of the voltage C at different areas (see FIG.~\ref{fig_14um_geom}(b) and (e)). \newline  
The two-interface Seebeck effect discussed in the supplemental material explains why the thermovoltage depends linear on the laser spot position. Thus, the two-interface Seebeck effect in the supplemental material explains the linear region of the contour plot in FIG.~\ref{fig_14um_geom}(a) and (d) but not the enhancement of the thermovoltage at the metal contacts. Hence, an additional effect is needed to explain the thermovoltage enhancement at the nanowire/contact interface. The band gap of $\text{Bi}_2\text{Se}_3$ is $300\,\mathrm{meV}$ which is typical for semiconductors. Therefore, we expect a band bending at the interface as sketched in FIG.~\ref{fig_14um_geom}(g). Due to the band bending, a positive (negative) space charge zone forms at the contact (nanowire) at the interface. When the laser illuminates the nanowire underneath the drain electrode, an electron-hole pair is created. The electron will be accelerated to the drain electrode and the hole to the center of the nanowire, driven by the electric field around the interface. The electrons entering the drain electrode will contribute to the net-positive current when the holes annihilate with electrons, originating from the source electrode. The photoelectrons generated close to the source electrode contribute to the net-negative current, since the electric field of the space charge region changes it sign. The enhancement at the drain electrode has a maximum at position (7,7) and decreases when the laser spot is moved away from this position. The decrease is due to the Gaussian laser intensity profile. When the laser spot is moved away from position (7,7), less photons illuminate the interface and the current contributing to the measured voltage decreases. The same effect appears at the source electrode, but with an opposite sign. \newline Comparing the position of the extremal voltage C in FIG.~\ref{fig_14um_geom}(e) to the device layout shows, that the largest and smallest spin-polarized current values are located at the edges of the contacts. This is similar to our earlier observations \cite{Thomas2018} on photocurrent measurements on $(\text{Bi}_{0.57}\text{Sb}_{0.43})_2\text{Te}_3$ Hall bars. There we report the accumulation of spin-polarized current at the Hall bar edges, with a different sign at opposite edges. We explained this by the spin Nernst effect, which describes the separation of electrons with opposite spin orientations due to a transverse temperature gradient. How the spin Nernst effect can enhance the values for C in the nanowire (NW) devices is depicted for the drain electrode in FIG.~\ref{fig_14um_geom}(h). When the laser spot illuminates the left edge of the drain electrode the electrons with opposite spin orientations are deflected perpendicular to the temperature gradient $\nabla T$. The spin-up electrons (depicted in blue) are deflected towards the interface and can enter the gold layer. Hence, the population of the Dirac cone is asymmetric, before taking optical transitions into account. At the right edge of the drain electrode, the temperature gradient $\nabla T$ points in the opposite direction. Thus, the spin-down electrons (depicted in red) are deflected into the gold layer. This means more surface states are populated by spin-up than spin-down electrons. Taking the optical excitation with left-circular polarized light into account, decreases (increases) the asymmetry of the surface state population at the left (right) edge of the drain electrode. For right-circular polarized light the asymmetry would increases (decreases) at the left (right) edge of the drain electrode. Since C is the photovoltage difference for left- and right-circular polarized light, it is positive (negative) at the left (right) side of the drain electrode. This leads to an increase of C at the contact edges on top of the nanowire (see FIG.~\ref{fig_14um_geom}(e) and (b)). At the source electrode, the temperature gradient is negative compared to the drain electrode (see FIG.~\ref{fig_14um_geom}(a)). Hence we expect a similar distribution of C with opposite sign, i.e. we expect C to be positive (negative) at the right side (left) of the source electrode in FIG.~\ref{fig_14um_geom}(e). \newline In summary, spatially resolved photocurrent measurements on $\text{Bi}_2\text{Se}_3$ nanowires were performed, focusing on the influence of the contact/nanowire interface on the polarization-independent and the spin-polarized current. Photocurrent measurements on several nanowires yield qualitatively similar results. The polarization-independent contribution on the contact/nanowire interface is a combination of the Seebeck effect  and the band bending at the contact/nanowire interface. The Seebeck effect explains the sign change of the thermovoltage  along the nanowire. The acceleration of the photoelectrons generated in the space charge region of the contact/nanowire interface causes the enhancement. The Gaussian profile of the laser spot leads to the radial symmetry of the enhancement located at the nanowire contact crossing. We observe for the spin-polarized current C (or voltage C generated by a spin-polarized current respectively) a constant region in the middle of the nanowire and an enhancement of the current at the contact edges together with a sign change of the spin-polarized current at each contact. The enhancement of the spin-polarized current close to the nanowire/metallic contact interface might be caused by the spin Nernst effect. We show that the influence of the gold contacts to the overall behavior of the nanowire is significant. The enhancement of the spin-polarized current, created in the vicinity of the contact pads, is in good agreement with the findings reported by Seifert et al. \cite{Seifert2017}. However, in our low excitation regime, the enhancement is caused by band-bending effects at the contact/nanowire interface and the spin Nernst effect. Enhancing spin-polarized currents in TI nanowires opens up further possibilities for spin current engineering, e.g. generating circular-polarized THz radiation by crossed nanowires.\newline Supplementary Material: See supplementary material for the discussion of the two-interface Seebeck
model for the nanowire devices.\newline The We are grateful to the German Science Foundation (DFG) for financial support through the priority program SPP1666: `Topological insulators: materials, fundamental properties, devices` (MU1780/10-2).\newline Data availability: The data that support the findings of this study are available from the corresponding author upon reasonable request.

This article may be downloaded for personal use only. Any other use requires prior permission of the author and AIP Publishing. This article appeared in Appl. Phys. Lett. \textbf{117}, 262401 (2020) and may be found at \url{https://doi.org/10.1063/5.0019044}.
\bibliography{references}

\providecommand{\noopsort}[1]{}\providecommand{\singleletter}[1]{#1}%
\begin{thebibliography}{25}%
\makeatletter
\providecommand \@ifxundefined [1]{%
 \@ifx{#1\undefined}
}%
\providecommand \@ifnum [1]{%
 \ifnum #1\expandafter \@firstoftwo
 \else \expandafter \@secondoftwo
 \fi
}%
\providecommand \@ifx [1]{%
 \ifx #1\expandafter \@firstoftwo
 \else \expandafter \@secondoftwo
 \fi
}%
\providecommand \natexlab [1]{#1}%
\providecommand \enquote  [1]{``#1''}%
\providecommand \bibnamefont  [1]{#1}%
\providecommand \bibfnamefont [1]{#1}%
\providecommand \citenamefont [1]{#1}%
\providecommand \href@noop [0]{\@secondoftwo}%
\providecommand \href [0]{\begingroup \@sanitize@url \@href}%
\providecommand \@href[1]{\@@startlink{#1}\@@href}%
\providecommand \@@href[1]{\endgroup#1\@@endlink}%
\providecommand \@sanitize@url [0]{\catcode `\\12\catcode `\$12\catcode
  `\&12\catcode `\#12\catcode `\^12\catcode `\_12\catcode `\%12\relax}%
\providecommand \@@startlink[1]{}%
\providecommand \@@endlink[0]{}%
\providecommand \url  [0]{\begingroup\@sanitize@url \@url }%
\providecommand \@url [1]{\endgroup\@href {#1}{\urlprefix }}%
\providecommand \urlprefix  [0]{URL }%
\providecommand \Eprint [0]{\href }%
\providecommand \doibase [0]{https://doi.org/}%
\providecommand \selectlanguage [0]{\@gobble}%
\providecommand \bibinfo  [0]{\@secondoftwo}%
\providecommand \bibfield  [0]{\@secondoftwo}%
\providecommand \translation [1]{[#1]}%
\providecommand \BibitemOpen [0]{}%
\providecommand \bibitemStop [0]{}%
\providecommand \bibitemNoStop [0]{.\EOS\space}%
\providecommand \EOS [0]{\spacefactor3000\relax}%
\providecommand \BibitemShut  [1]{\csname bibitem#1\endcsname}%
\let\auto@bib@innerbib\@empty
\bibitem [{\citenamefont {Thouless}\ \emph {et~al.}(1982)\citenamefont
  {Thouless}, \citenamefont {Kohmoto}, \citenamefont {Nightingale},\ and\
  \citenamefont {den Nijs}}]{Thouless1982}%
  \BibitemOpen
  \bibfield  {author} {\bibinfo {author} {\bibfnamefont {D.~J.}\ \bibnamefont
  {Thouless}}, \bibinfo {author} {\bibfnamefont {M.}~\bibnamefont {Kohmoto}},
  \bibinfo {author} {\bibfnamefont {M.~P.}\ \bibnamefont {Nightingale}},\ and\
  \bibinfo {author} {\bibfnamefont {M.}~\bibnamefont {den Nijs}},\ }\bibfield
  {title} {\enquote {\bibinfo {title} {{Q}uantized {H}all {C}onductance in a
  {T}wo-{D}imensional {P}eriodic {P}otential},}\ }\href
  {https://doi.org/10.1103/PhysRevLett.49.405} {\bibfield  {journal} {\bibinfo
  {journal} {Phys. Rev. Lett.}\ }\textbf {\bibinfo {volume} {49}},\ \bibinfo
  {pages} {405--408} (\bibinfo {year} {1982})}\BibitemShut {NoStop}%
\bibitem [{\citenamefont {Kane}\ and\ \citenamefont
  {Mele}(2005)}]{KaneMele2005}%
  \BibitemOpen
  \bibfield  {author} {\bibinfo {author} {\bibfnamefont {C.~L.}\ \bibnamefont
  {Kane}}\ and\ \bibinfo {author} {\bibfnamefont {E.~J.}\ \bibnamefont
  {Mele}},\ }\bibfield  {title} {\enquote {\bibinfo {title} {${Z}_{2}$
  {T}opological {O}rder and the {Q}uantum {S}pin {H}all {E}ffect},}\ }\href
  {https://doi.org/10.1103/PhysRevLett.95.146802} {\bibfield  {journal}
  {\bibinfo  {journal} {Phys. Rev. Lett.}\ }\textbf {\bibinfo {volume} {95}},\
  \bibinfo {pages} {146802} (\bibinfo {year} {2005})}\BibitemShut {NoStop}%
\bibitem [{\citenamefont {K{\"o}nig}\ \emph {et~al.}(2007)\citenamefont
  {K{\"o}nig}, \citenamefont {Wiedmann}, \citenamefont {Br{\"u}ne},
  \citenamefont {Roth}, \citenamefont {Buhmann}, \citenamefont {Molenkamp},
  \citenamefont {Qi},\ and\ \citenamefont {Zhang}}]{Koenig2007}%
  \BibitemOpen
  \bibfield  {author} {\bibinfo {author} {\bibfnamefont {M.}~\bibnamefont
  {K{\"o}nig}}, \bibinfo {author} {\bibfnamefont {S.}~\bibnamefont {Wiedmann}},
  \bibinfo {author} {\bibfnamefont {C.}~\bibnamefont {Br{\"u}ne}}, \bibinfo
  {author} {\bibfnamefont {A.}~\bibnamefont {Roth}}, \bibinfo {author}
  {\bibfnamefont {H.}~\bibnamefont {Buhmann}}, \bibinfo {author} {\bibfnamefont
  {L.~W.}\ \bibnamefont {Molenkamp}}, \bibinfo {author} {\bibfnamefont {X.-L.}\
  \bibnamefont {Qi}},\ and\ \bibinfo {author} {\bibfnamefont {S.-C.}\
  \bibnamefont {Zhang}},\ }\bibfield  {title} {\enquote {\bibinfo {title}
  {{Q}uantum {S}pin {H}all {I}nsulator {S}tate in {H}g{T}e {Q}uantum
  {W}ells},}\ }\href {https://doi.org/10.1126/science.1148047} {\bibfield
  {journal} {\bibinfo  {journal} {Science}\ }\textbf {\bibinfo {volume}
  {318}},\ \bibinfo {pages} {766--770} (\bibinfo {year} {2007})}\BibitemShut
  {NoStop}%
\bibitem [{\citenamefont {Zhang}\ \emph {et~al.}(2009)\citenamefont {Zhang},
  \citenamefont {Liu}, \citenamefont {Qi}, \citenamefont {Dai}, \citenamefont
  {Fang},\ and\ \citenamefont {Zhang}}]{Zhang2009}%
  \BibitemOpen
  \bibfield  {author} {\bibinfo {author} {\bibfnamefont {H.}~\bibnamefont
  {Zhang}}, \bibinfo {author} {\bibfnamefont {C.-X.}\ \bibnamefont {Liu}},
  \bibinfo {author} {\bibfnamefont {X.-L.}\ \bibnamefont {Qi}}, \bibinfo
  {author} {\bibfnamefont {X.}~\bibnamefont {Dai}}, \bibinfo {author}
  {\bibfnamefont {Z.}~\bibnamefont {Fang}},\ and\ \bibinfo {author}
  {\bibfnamefont {S.-C.}\ \bibnamefont {Zhang}},\ }\bibfield  {title} {\enquote
  {\bibinfo {title} {{T}opological insulators in $\text{Bi}_2\text{Se}_3$,
  $\text{Bi}_2\text{Te}_3$ and $\text{Sb}_2\text{Te}_3$ with a single {D}irac
  cone on the surface},}\ }\href {https://doi.org/10.1038/nphys1270} {\bibfield
   {journal} {\bibinfo  {journal} {Nature Physics}\ }\textbf {\bibinfo {volume}
  {5}},\ \bibinfo {pages} {438--442} (\bibinfo {year} {2009})}\BibitemShut
  {NoStop}%
\bibitem [{\citenamefont {Xia}\ \emph {et~al.}(2009)\citenamefont {Xia},
  \citenamefont {Qian}, \citenamefont {Hsieh}, \citenamefont {Wray},
  \citenamefont {Pal}, \citenamefont {Lin}, \citenamefont {Bansil},
  \citenamefont {Grauer}, \citenamefont {Hor}, \citenamefont {Cava},\ and\
  \citenamefont {Hasan}}]{Xia2009}%
  \BibitemOpen
  \bibfield  {author} {\bibinfo {author} {\bibfnamefont {Y.}~\bibnamefont
  {Xia}}, \bibinfo {author} {\bibfnamefont {D.}~\bibnamefont {Qian}}, \bibinfo
  {author} {\bibfnamefont {D.}~\bibnamefont {Hsieh}}, \bibinfo {author}
  {\bibfnamefont {L.}~\bibnamefont {Wray}}, \bibinfo {author} {\bibfnamefont
  {A.}~\bibnamefont {Pal}}, \bibinfo {author} {\bibfnamefont {H.}~\bibnamefont
  {Lin}}, \bibinfo {author} {\bibfnamefont {A.}~\bibnamefont {Bansil}},
  \bibinfo {author} {\bibfnamefont {D.}~\bibnamefont {Grauer}}, \bibinfo
  {author} {\bibfnamefont {Y.~S.}\ \bibnamefont {Hor}}, \bibinfo {author}
  {\bibfnamefont {R.~J.}\ \bibnamefont {Cava}},\ and\ \bibinfo {author}
  {\bibfnamefont {M.~Z.}\ \bibnamefont {Hasan}},\ }\bibfield  {title} {\enquote
  {\bibinfo {title} {Observation of a large-gap topological-insulator class
  with a single {D}irac cone on the surface},}\ }\href
  {https://doi.org/10.1038/nphys1274} {\bibfield  {journal} {\bibinfo
  {journal} {Nature Physics}\ }\textbf {\bibinfo {volume} {5}},\ \bibinfo
  {pages} {398--402} (\bibinfo {year} {2009})}\BibitemShut {NoStop}%
\bibitem [{\citenamefont {Hasan}\ and\ \citenamefont
  {Kane}(2010)}]{HasanKane2010}%
  \BibitemOpen
  \bibfield  {author} {\bibinfo {author} {\bibfnamefont {M.~Z.}\ \bibnamefont
  {Hasan}}\ and\ \bibinfo {author} {\bibfnamefont {C.~L.}\ \bibnamefont
  {Kane}},\ }\bibfield  {title} {\enquote {\bibinfo {title} {Colloquium:
  Topological insulators},}\ }\href
  {https://doi.org/10.1103/RevModPhys.82.3045} {\bibfield  {journal} {\bibinfo
  {journal} {Rev. Mod. Phys.}\ }\textbf {\bibinfo {volume} {82}},\ \bibinfo
  {pages} {3045--3067} (\bibinfo {year} {2010})}\BibitemShut {NoStop}%
\bibitem [{\citenamefont {Hsieh}\ \emph {et~al.}(2008)\citenamefont {Hsieh},
  \citenamefont {Qian}, \citenamefont {Wray}, \citenamefont {Xia},
  \citenamefont {Hor}, \citenamefont {Cava},\ and\ \citenamefont
  {Hasan}}]{Hsieh2008}%
  \BibitemOpen
  \bibfield  {author} {\bibinfo {author} {\bibfnamefont {D.}~\bibnamefont
  {Hsieh}}, \bibinfo {author} {\bibfnamefont {D.}~\bibnamefont {Qian}},
  \bibinfo {author} {\bibfnamefont {L.}~\bibnamefont {Wray}}, \bibinfo {author}
  {\bibfnamefont {Y.}~\bibnamefont {Xia}}, \bibinfo {author} {\bibfnamefont
  {Y.~S.}\ \bibnamefont {Hor}}, \bibinfo {author} {\bibfnamefont {R.~J.}\
  \bibnamefont {Cava}},\ and\ \bibinfo {author} {\bibfnamefont {M.~Z.}\
  \bibnamefont {Hasan}},\ }\bibfield  {title} {\enquote {\bibinfo {title} {A
  topological {D}irac insulator in a quantum spin {H}all phase},}\ }\href
  {https://doi.org/10.1038/nature06843} {\bibfield  {journal} {\bibinfo
  {journal} {Nature}\ }\textbf {\bibinfo {volume} {452}},\ \bibinfo {pages}
  {970--974} (\bibinfo {year} {2008})}\BibitemShut {NoStop}%
\bibitem [{\citenamefont {Yan}\ \emph {et~al.}(2014)\citenamefont {Yan},
  \citenamefont {Liao}, \citenamefont {Ke}, \citenamefont {Van~Tendeloo},
  \citenamefont {Wang}, \citenamefont {Sun}, \citenamefont {Yao}, \citenamefont
  {Zhou}, \citenamefont {Zhang}, \citenamefont {Wu},\ and\ \citenamefont
  {Yu}}]{Yan2014}%
  \BibitemOpen
  \bibfield  {author} {\bibinfo {author} {\bibfnamefont {Y.}~\bibnamefont
  {Yan}}, \bibinfo {author} {\bibfnamefont {Z.-M.}\ \bibnamefont {Liao}},
  \bibinfo {author} {\bibfnamefont {X.}~\bibnamefont {Ke}}, \bibinfo {author}
  {\bibfnamefont {G.}~\bibnamefont {Van~Tendeloo}}, \bibinfo {author}
  {\bibfnamefont {Q.}~\bibnamefont {Wang}}, \bibinfo {author} {\bibfnamefont
  {D.}~\bibnamefont {Sun}}, \bibinfo {author} {\bibfnamefont {W.}~\bibnamefont
  {Yao}}, \bibinfo {author} {\bibfnamefont {S.}~\bibnamefont {Zhou}}, \bibinfo
  {author} {\bibfnamefont {L.}~\bibnamefont {Zhang}}, \bibinfo {author}
  {\bibfnamefont {H.-C.}\ \bibnamefont {Wu}},\ and\ \bibinfo {author}
  {\bibfnamefont {D.-P.}\ \bibnamefont {Yu}},\ }\bibfield  {title} {\enquote
  {\bibinfo {title} {{T}opological {S}urface {S}tate {E}nhanced
  {P}hotothermoelectric {E}ffect in $\text{Bi}_2\text{Se}_3$ {N}anoribbons},}\
  }\href {https://doi.org/10.1021/nl501276e} {\bibfield  {journal} {\bibinfo
  {journal} {Nano Letters}\ }\textbf {\bibinfo {volume} {14}},\ \bibinfo
  {pages} {4389--4394} (\bibinfo {year} {2014})}\BibitemShut {NoStop}%
\bibitem [{\citenamefont {Sharma}\ \emph {et~al.}(2016)\citenamefont {Sharma},
  \citenamefont {Bhattacharyya}, \citenamefont {Srivastava}, \citenamefont
  {Senguttuvan},\ and\ \citenamefont {Husale}}]{Sharma2016}%
  \BibitemOpen
  \bibfield  {author} {\bibinfo {author} {\bibfnamefont {A.}~\bibnamefont
  {Sharma}}, \bibinfo {author} {\bibfnamefont {B.}~\bibnamefont
  {Bhattacharyya}}, \bibinfo {author} {\bibfnamefont {A.~K.}\ \bibnamefont
  {Srivastava}}, \bibinfo {author} {\bibfnamefont {T.~D.}\ \bibnamefont
  {Senguttuvan}},\ and\ \bibinfo {author} {\bibfnamefont {S.}~\bibnamefont
  {Husale}},\ }\bibfield  {title} {\enquote {\bibinfo {title} {High performance
  broadband photodetector using fabricated nanowires of bismuth selenide},}\
  }\href {https://doi.org/10.1038/srep19138} {\bibfield  {journal} {\bibinfo
  {journal} {Scientific Reports}\ }\textbf {\bibinfo {volume} {6}},\ \bibinfo
  {pages} {19138} (\bibinfo {year} {2016})}\BibitemShut {NoStop}%
\bibitem [{\citenamefont {Ganichev}\ and\ \citenamefont
  {Prettl}(2003)}]{Ganichev2003}%
  \BibitemOpen
  \bibfield  {author} {\bibinfo {author} {\bibfnamefont {S.~D.}\ \bibnamefont
  {Ganichev}}\ and\ \bibinfo {author} {\bibfnamefont {W.}~\bibnamefont
  {Prettl}},\ }\bibfield  {title} {\enquote {\bibinfo {title} {Spin
  photocurrents in quantum wells},}\ }\href
  {https://doi.org/10.1088/0953-8984/15/20/204} {\bibfield  {journal} {\bibinfo
   {journal} {Journal of Physics: Condensed Matter}\ }\textbf {\bibinfo
  {volume} {15}},\ \bibinfo {pages} {R935--R983} (\bibinfo {year}
  {2003})}\BibitemShut {NoStop}%
\bibitem [{\citenamefont {McIver}\ \emph {et~al.}(2011)\citenamefont {McIver},
  \citenamefont {Hsieh}, \citenamefont {Steinberg}, \citenamefont
  {Jarillo-Herrero},\ and\ \citenamefont {Gedik}}]{McIver2011}%
  \BibitemOpen
  \bibfield  {author} {\bibinfo {author} {\bibfnamefont {J.~W.}\ \bibnamefont
  {McIver}}, \bibinfo {author} {\bibfnamefont {D.~C.}\ \bibnamefont {Hsieh}},
  \bibinfo {author} {\bibfnamefont {H.}~\bibnamefont {Steinberg}}, \bibinfo
  {author} {\bibfnamefont {P.}~\bibnamefont {Jarillo-Herrero}},\ and\ \bibinfo
  {author} {\bibfnamefont {N.}~\bibnamefont {Gedik}},\ }\bibfield  {title}
  {\enquote {\bibinfo {title} {Control over topological insulator photocurrents
  with light polarization.}}\ }\href@noop {} {\bibfield  {journal} {\bibinfo
  {journal} {Nature nanotechnology}\ }\textbf {\bibinfo {volume} {72}},\
  \bibinfo {pages} {96--100} (\bibinfo {year} {2011})}\BibitemShut {NoStop}%
\bibitem [{\citenamefont {Kastl}\ \emph {et~al.}(2015)\citenamefont {Kastl},
  \citenamefont {Karnetzky}, \citenamefont {Karl},\ and\ \citenamefont
  {Holleitner}}]{Kastl2015}%
  \BibitemOpen
  \bibfield  {author} {\bibinfo {author} {\bibfnamefont {C.}~\bibnamefont
  {Kastl}}, \bibinfo {author} {\bibfnamefont {C.}~\bibnamefont {Karnetzky}},
  \bibinfo {author} {\bibfnamefont {H.}~\bibnamefont {Karl}},\ and\ \bibinfo
  {author} {\bibfnamefont {A.~W.}\ \bibnamefont {Holleitner}},\ }\bibfield
  {title} {\enquote {\bibinfo {title} {Ultrafast helicity control of surface
  currents in topological insulators with near-unity fidelity},}\ }\href
  {https://doi.org/10.1038/ncomms7617} {\bibfield  {journal} {\bibinfo
  {journal} {Nature Communications}\ }\textbf {\bibinfo {volume} {6}},\
  \bibinfo {pages} {6617} (\bibinfo {year} {2015})}\BibitemShut {NoStop}%
\bibitem [{\citenamefont {Seifert}\ \emph {et~al.}(2017)\citenamefont
  {Seifert}, \citenamefont {Vaklinova}, \citenamefont {Kern}, \citenamefont
  {Burghard},\ and\ \citenamefont {Holleitner}}]{Seifert2017}%
  \BibitemOpen
  \bibfield  {author} {\bibinfo {author} {\bibfnamefont {P.}~\bibnamefont
  {Seifert}}, \bibinfo {author} {\bibfnamefont {K.}~\bibnamefont {Vaklinova}},
  \bibinfo {author} {\bibfnamefont {K.}~\bibnamefont {Kern}}, \bibinfo {author}
  {\bibfnamefont {M.}~\bibnamefont {Burghard}},\ and\ \bibinfo {author}
  {\bibfnamefont {A.}~\bibnamefont {Holleitner}},\ }\bibfield  {title}
  {\enquote {\bibinfo {title} {{S}urface {S}tate-{D}ominated photoconduction
  and {TH}z {G}eneration in {T}opological $\text{Bi}_2\text{Te}_2\text{Se}$
  {N}anowires},}\ }\href {https://doi.org/10.1021/acs.nanolett.6b04312}
  {\bibfield  {journal} {\bibinfo  {journal} {Nano Letters}\ }\textbf {\bibinfo
  {volume} {17}},\ \bibinfo {pages} {973--979} (\bibinfo {year}
  {2017})}\BibitemShut {NoStop}%
\bibitem [{\citenamefont {Meyer}\ \emph {et~al.}(2020)\citenamefont {Meyer},
  \citenamefont {Geishendorf}, \citenamefont {Walowski}, \citenamefont
  {Thomas},\ and\ \citenamefont {M{\"u}nzenberg}}]{Meyer2020}%
  \BibitemOpen
  \bibfield  {author} {\bibinfo {author} {\bibfnamefont {N.}~\bibnamefont
  {Meyer}}, \bibinfo {author} {\bibfnamefont {K.}~\bibnamefont {Geishendorf}},
  \bibinfo {author} {\bibfnamefont {J.}~\bibnamefont {Walowski}}, \bibinfo
  {author} {\bibfnamefont {A.}~\bibnamefont {Thomas}},\ and\ \bibinfo {author}
  {\bibfnamefont {M.}~\bibnamefont {M{\"u}nzenberg}},\ }\bibfield  {title}
  {\enquote {\bibinfo {title} {{P}hotocurrent measurements in topological
  insulator $\text{Bi}_2\text{Se}_3$ nanowires},}\ }\href
  {https://doi.org/10.1063/1.5142837} {\bibfield  {journal} {\bibinfo
  {journal} {Applied Physics Letters}\ }\textbf {\bibinfo {volume} {116}},\
  \bibinfo {pages} {172402} (\bibinfo {year} {2020})},\ \Eprint
  {https://arxiv.org/abs/https://doi.org/10.1063/1.5142837}
  {https://doi.org/10.1063/1.5142837} \BibitemShut {NoStop}%
\bibitem [{\citenamefont {Shin}\ \emph {et~al.}(2016)\citenamefont {Shin},
  \citenamefont {Hamdou}, \citenamefont {Reith}, \citenamefont {Osterhage},
  \citenamefont {Gooth}, \citenamefont {Damm}, \citenamefont {Rellinghaus},
  \citenamefont {Pippel},\ and\ \citenamefont {Nielsch}}]{Shin2016}%
  \BibitemOpen
  \bibfield  {author} {\bibinfo {author} {\bibfnamefont {H.~S.}\ \bibnamefont
  {Shin}}, \bibinfo {author} {\bibfnamefont {B.}~\bibnamefont {Hamdou}},
  \bibinfo {author} {\bibfnamefont {H.}~\bibnamefont {Reith}}, \bibinfo
  {author} {\bibfnamefont {H.}~\bibnamefont {Osterhage}}, \bibinfo {author}
  {\bibfnamefont {J.}~\bibnamefont {Gooth}}, \bibinfo {author} {\bibfnamefont
  {C.}~\bibnamefont {Damm}}, \bibinfo {author} {\bibfnamefont {B.}~\bibnamefont
  {Rellinghaus}}, \bibinfo {author} {\bibfnamefont {E.}~\bibnamefont
  {Pippel}},\ and\ \bibinfo {author} {\bibfnamefont {K.}~\bibnamefont
  {Nielsch}},\ }\bibfield  {title} {\enquote {\bibinfo {title} {{T}he
  surface-to-volume ratio: a key parameter in the thermoelectric transport of
  topological insulator $\text{Bi}_2\text{Se}_3$ nanowires},}\ }\href
  {https://doi.org/10.1039/C6NR01716A} {\bibfield  {journal} {\bibinfo
  {journal} {Nanoscale}\ }\textbf {\bibinfo {volume} {8}},\ \bibinfo {pages}
  {13552--13557} (\bibinfo {year} {2016})}\BibitemShut {NoStop}%
\bibitem [{\citenamefont {Gooth}\ \emph {et~al.}(2014)\citenamefont {Gooth},
  \citenamefont {Gluschke}, \citenamefont {Zierold}, \citenamefont {Leijnse},
  \citenamefont {Linke},\ and\ \citenamefont {Nielsch}}]{Gooth2014}%
  \BibitemOpen
  \bibfield  {author} {\bibinfo {author} {\bibfnamefont {J.}~\bibnamefont
  {Gooth}}, \bibinfo {author} {\bibfnamefont {J.~G.}\ \bibnamefont {Gluschke}},
  \bibinfo {author} {\bibfnamefont {R.}~\bibnamefont {Zierold}}, \bibinfo
  {author} {\bibfnamefont {M.}~\bibnamefont {Leijnse}}, \bibinfo {author}
  {\bibfnamefont {H.}~\bibnamefont {Linke}},\ and\ \bibinfo {author}
  {\bibfnamefont {K.}~\bibnamefont {Nielsch}},\ }\bibfield  {title} {\enquote
  {\bibinfo {title} {Thermoelectric performance of classical topological
  insulator nanowires},}\ }\href
  {https://doi.org/10.1088/0268-1242/30/1/015015} {\bibfield  {journal}
  {\bibinfo  {journal} {Semiconductor Science and Technology}\ }\textbf
  {\bibinfo {volume} {30}},\ \bibinfo {pages} {015015} (\bibinfo {year}
  {2014})}\BibitemShut {NoStop}%
\bibitem [{\citenamefont {Shalygin}\ \emph {et~al.}(2007)\citenamefont
  {Shalygin}, \citenamefont {Diehl}, \citenamefont {Hoffmann}, \citenamefont
  {Danilov}, \citenamefont {Herrle}, \citenamefont {Tarasenko}, \citenamefont
  {Schuh}, \citenamefont {Gerl}, \citenamefont {Wegscheider}, \citenamefont
  {Prettl},\ and\ \citenamefont {Ganichev}}]{Shalygin2007}%
  \BibitemOpen
  \bibfield  {author} {\bibinfo {author} {\bibfnamefont {V.~A.}\ \bibnamefont
  {Shalygin}}, \bibinfo {author} {\bibfnamefont {H.}~\bibnamefont {Diehl}},
  \bibinfo {author} {\bibfnamefont {C.}~\bibnamefont {Hoffmann}}, \bibinfo
  {author} {\bibfnamefont {S.~N.}\ \bibnamefont {Danilov}}, \bibinfo {author}
  {\bibfnamefont {T.}~\bibnamefont {Herrle}}, \bibinfo {author} {\bibfnamefont
  {S.~A.}\ \bibnamefont {Tarasenko}}, \bibinfo {author} {\bibfnamefont
  {D.}~\bibnamefont {Schuh}}, \bibinfo {author} {\bibfnamefont
  {C.}~\bibnamefont {Gerl}}, \bibinfo {author} {\bibfnamefont {W.}~\bibnamefont
  {Wegscheider}}, \bibinfo {author} {\bibfnamefont {W.}~\bibnamefont
  {Prettl}},\ and\ \bibinfo {author} {\bibfnamefont {S.~D.}\ \bibnamefont
  {Ganichev}},\ }\bibfield  {title} {\enquote {\bibinfo {title} {Spin
  photocurrents and the circular photon drag effect in (110)-grown quantum well
  structures},}\ }\href {https://doi.org/10.1134/S0021364006220097} {\bibfield
  {journal} {\bibinfo  {journal} {JETP Letters}\ }\textbf {\bibinfo {volume}
  {84}},\ \bibinfo {pages} {570--576} (\bibinfo {year} {2007})}\BibitemShut
  {NoStop}%
\bibitem [{\citenamefont {Takeno}, \citenamefont {Saito},\ and\ \citenamefont
  {Mizoguchi}(2018)}]{Takeno2018}%
  \BibitemOpen
  \bibfield  {author} {\bibinfo {author} {\bibfnamefont {H.}~\bibnamefont
  {Takeno}}, \bibinfo {author} {\bibfnamefont {S.}~\bibnamefont {Saito}},\ and\
  \bibinfo {author} {\bibfnamefont {K.}~\bibnamefont {Mizoguchi}},\ }\bibfield
  {title} {\enquote {\bibinfo {title} {Optical control of spin-polarized
  photocurrent in topological insulator thin films},}\ }\href
  {https://doi.org/10.1038/s41598-018-33716-0} {\bibfield  {journal} {\bibinfo
  {journal} {Scientific Reports}\ }\textbf {\bibinfo {volume} {8}},\ \bibinfo
  {pages} {15392} (\bibinfo {year} {2018})}\BibitemShut {NoStop}%
\bibitem [{\citenamefont {Okada}\ \emph {et~al.}(2016)\citenamefont {Okada},
  \citenamefont {Ogawa}, \citenamefont {Yoshimi}, \citenamefont {Tsukazaki},
  \citenamefont {Takahashi}, \citenamefont {Kawasaki},\ and\ \citenamefont
  {Tokura}}]{Okada2016}%
  \BibitemOpen
  \bibfield  {author} {\bibinfo {author} {\bibfnamefont {K.~N.}\ \bibnamefont
  {Okada}}, \bibinfo {author} {\bibfnamefont {N.}~\bibnamefont {Ogawa}},
  \bibinfo {author} {\bibfnamefont {R.}~\bibnamefont {Yoshimi}}, \bibinfo
  {author} {\bibfnamefont {A.}~\bibnamefont {Tsukazaki}}, \bibinfo {author}
  {\bibfnamefont {K.~S.}\ \bibnamefont {Takahashi}}, \bibinfo {author}
  {\bibfnamefont {M.}~\bibnamefont {Kawasaki}},\ and\ \bibinfo {author}
  {\bibfnamefont {Y.}~\bibnamefont {Tokura}},\ }\bibfield  {title} {\enquote
  {\bibinfo {title} {Enhanced photogalvanic current in topological insulators
  via fermi energy tuning},}\ }\href
  {https://doi.org/10.1103/PhysRevB.93.081403} {\bibfield  {journal} {\bibinfo
  {journal} {Phys. Rev. B}\ }\textbf {\bibinfo {volume} {93}},\ \bibinfo
  {pages} {081403} (\bibinfo {year} {2016})}\BibitemShut {NoStop}%
\bibitem [{\citenamefont {Plank}\ \emph
  {et~al.}(2016{\natexlab{a}})\citenamefont {Plank}, \citenamefont {Danilov},
  \citenamefont {Bel'kov}, \citenamefont {Shalygin}, \citenamefont {Kampmeier},
  \citenamefont {Lanius}, \citenamefont {Mussler}, \citenamefont
  {Gr{\"u}tzmacher},\ and\ \citenamefont {Ganichev}}]{Plank2016}%
  \BibitemOpen
  \bibfield  {author} {\bibinfo {author} {\bibfnamefont {H.}~\bibnamefont
  {Plank}}, \bibinfo {author} {\bibfnamefont {S.~N.}\ \bibnamefont {Danilov}},
  \bibinfo {author} {\bibfnamefont {V.~V.}\ \bibnamefont {Bel'kov}}, \bibinfo
  {author} {\bibfnamefont {V.~A.}\ \bibnamefont {Shalygin}}, \bibinfo {author}
  {\bibfnamefont {J.}~\bibnamefont {Kampmeier}}, \bibinfo {author}
  {\bibfnamefont {M.}~\bibnamefont {Lanius}}, \bibinfo {author} {\bibfnamefont
  {G.}~\bibnamefont {Mussler}}, \bibinfo {author} {\bibfnamefont
  {D.}~\bibnamefont {Gr{\"u}tzmacher}},\ and\ \bibinfo {author} {\bibfnamefont
  {S.~D.}\ \bibnamefont {Ganichev}},\ }\bibfield  {title} {\enquote {\bibinfo
  {title} {Opto-electronic characterization of three dimensional topological
  insulators},}\ }\href {https://doi.org/10.1063/1.4965962} {\bibfield
  {journal} {\bibinfo  {journal} {Journal of Applied Physics}\ }\textbf
  {\bibinfo {volume} {120}},\ \bibinfo {pages} {165301} (\bibinfo {year}
  {2016}{\natexlab{a}})}\BibitemShut {NoStop}%
\bibitem [{\citenamefont {{Braun}}\ \emph {et~al.}(2016)\citenamefont
  {{Braun}}, \citenamefont {{Mussler}}, \citenamefont {{Hruban}}, \citenamefont
  {{Konczykowski}}, \citenamefont {{Schumann}}, \citenamefont {{Wolf}},
  \citenamefont {{M{\"u}nzenberg}}, \citenamefont {{Perfetti}},\ and\
  \citenamefont {{Kampfrath}}}]{Braun2016}%
  \BibitemOpen
  \bibfield  {author} {\bibinfo {author} {\bibfnamefont {L.}~\bibnamefont
  {{Braun}}}, \bibinfo {author} {\bibfnamefont {G.}~\bibnamefont {{Mussler}}},
  \bibinfo {author} {\bibfnamefont {A.}~\bibnamefont {{Hruban}}}, \bibinfo
  {author} {\bibfnamefont {M.}~\bibnamefont {{Konczykowski}}}, \bibinfo
  {author} {\bibfnamefont {T.}~\bibnamefont {{Schumann}}}, \bibinfo {author}
  {\bibfnamefont {M.}~\bibnamefont {{Wolf}}}, \bibinfo {author} {\bibfnamefont
  {M.}~\bibnamefont {{M{\"u}nzenberg}}}, \bibinfo {author} {\bibfnamefont
  {L.}~\bibnamefont {{Perfetti}}},\ and\ \bibinfo {author} {\bibfnamefont
  {T.}~\bibnamefont {{Kampfrath}}},\ }\bibfield  {title} {\enquote {\bibinfo
  {title} {{{U}ltrafast photocurrents at the surface of the three-dimensional
  topological insulator {B}i$_2${S}e$_3$}},}\ }\href
  {https://doi.org/10.1038/ncomms13259} {\bibfield  {journal} {\bibinfo
  {journal} {{Nature Communications}}\ }\textbf {\bibinfo {volume} {7}},\
  \bibinfo {pages} {2041--1723} (\bibinfo {year} {2016})}\BibitemShut {NoStop}%
\bibitem [{\citenamefont {Luo}, \citenamefont {He},\ and\ \citenamefont
  {Li}(2017)}]{Luo2017}%
  \BibitemOpen
  \bibfield  {author} {\bibinfo {author} {\bibfnamefont {S.}~\bibnamefont
  {Luo}}, \bibinfo {author} {\bibfnamefont {L.}~\bibnamefont {He}},\ and\
  \bibinfo {author} {\bibfnamefont {M.}~\bibnamefont {Li}},\ }\bibfield
  {title} {\enquote {\bibinfo {title} {Spin-momentum locked interaction between
  guided photons and surface electrons in topological insulators},}\ }\href
  {https://doi.org/10.1038/s41467-017-02264-y} {\bibfield  {journal} {\bibinfo
  {journal} {Nature Communications}\ }\textbf {\bibinfo {volume} {8}},\
  \bibinfo {pages} {2141} (\bibinfo {year} {2017})}\BibitemShut {NoStop}%
\bibitem [{\citenamefont {Olbrich}\ \emph {et~al.}(2014)\citenamefont
  {Olbrich}, \citenamefont {Golub}, \citenamefont {Herrmann}, \citenamefont
  {Danilov}, \citenamefont {Plank}, \citenamefont {Bel'kov}, \citenamefont
  {Mussler}, \citenamefont {Weyrich}, \citenamefont {Schneider}, \citenamefont
  {Kampmeier}, \citenamefont {Gr\"utzmacher}, \citenamefont {Plucinski},
  \citenamefont {Eschbach},\ and\ \citenamefont {Ganichev}}]{Olbrich2014}%
  \BibitemOpen
  \bibfield  {author} {\bibinfo {author} {\bibfnamefont {P.}~\bibnamefont
  {Olbrich}}, \bibinfo {author} {\bibfnamefont {L.~E.}\ \bibnamefont {Golub}},
  \bibinfo {author} {\bibfnamefont {T.}~\bibnamefont {Herrmann}}, \bibinfo
  {author} {\bibfnamefont {S.~N.}\ \bibnamefont {Danilov}}, \bibinfo {author}
  {\bibfnamefont {H.}~\bibnamefont {Plank}}, \bibinfo {author} {\bibfnamefont
  {V.~V.}\ \bibnamefont {Bel'kov}}, \bibinfo {author} {\bibfnamefont
  {G.}~\bibnamefont {Mussler}}, \bibinfo {author} {\bibfnamefont
  {C.}~\bibnamefont {Weyrich}}, \bibinfo {author} {\bibfnamefont {C.~M.}\
  \bibnamefont {Schneider}}, \bibinfo {author} {\bibfnamefont {J.}~\bibnamefont
  {Kampmeier}}, \bibinfo {author} {\bibfnamefont {D.}~\bibnamefont
  {Gr\"utzmacher}}, \bibinfo {author} {\bibfnamefont {L.}~\bibnamefont
  {Plucinski}}, \bibinfo {author} {\bibfnamefont {M.}~\bibnamefont
  {Eschbach}},\ and\ \bibinfo {author} {\bibfnamefont {S.~D.}\ \bibnamefont
  {Ganichev}},\ }\bibfield  {title} {\enquote {\bibinfo {title}
  {{R}oom-temperature high-frequency transport of {D}irac fermions in
  epitaxially grown $\text{Sb}_2\text{Te}_3-$ and $\text{Bi}_2\text{Te}_3-$
  based topological insulators},}\ }\href
  {https://doi.org/10.1103/PhysRevLett.113.096601} {\bibfield  {journal}
  {\bibinfo  {journal} {Phys. Rev. Lett.}\ }\textbf {\bibinfo {volume} {113}},\
  \bibinfo {pages} {096601} (\bibinfo {year} {2014})}\BibitemShut {NoStop}%
\bibitem [{\citenamefont {Plank}\ \emph
  {et~al.}(2016{\natexlab{b}})\citenamefont {Plank}, \citenamefont {Golub},
  \citenamefont {Bauer}, \citenamefont {Bel'kov}, \citenamefont {Herrmann},
  \citenamefont {Olbrich}, \citenamefont {Eschbach}, \citenamefont {Plucinski},
  \citenamefont {Schneider}, \citenamefont {Kampmeier}, \citenamefont {Lanius},
  \citenamefont {Mussler}, \citenamefont {Gr\"utzmacher},\ and\ \citenamefont
  {Ganichev}}]{Plank2016B}%
  \BibitemOpen
  \bibfield  {author} {\bibinfo {author} {\bibfnamefont {H.}~\bibnamefont
  {Plank}}, \bibinfo {author} {\bibfnamefont {L.~E.}\ \bibnamefont {Golub}},
  \bibinfo {author} {\bibfnamefont {S.}~\bibnamefont {Bauer}}, \bibinfo
  {author} {\bibfnamefont {V.~V.}\ \bibnamefont {Bel'kov}}, \bibinfo {author}
  {\bibfnamefont {T.}~\bibnamefont {Herrmann}}, \bibinfo {author}
  {\bibfnamefont {P.}~\bibnamefont {Olbrich}}, \bibinfo {author} {\bibfnamefont
  {M.}~\bibnamefont {Eschbach}}, \bibinfo {author} {\bibfnamefont
  {L.}~\bibnamefont {Plucinski}}, \bibinfo {author} {\bibfnamefont {C.~M.}\
  \bibnamefont {Schneider}}, \bibinfo {author} {\bibfnamefont {J.}~\bibnamefont
  {Kampmeier}}, \bibinfo {author} {\bibfnamefont {M.}~\bibnamefont {Lanius}},
  \bibinfo {author} {\bibfnamefont {G.}~\bibnamefont {Mussler}}, \bibinfo
  {author} {\bibfnamefont {D.}~\bibnamefont {Gr\"utzmacher}},\ and\ \bibinfo
  {author} {\bibfnamefont {S.~D.}\ \bibnamefont {Ganichev}},\ }\bibfield
  {title} {\enquote {\bibinfo {title} {Photon drag effect in
  ($\text{Bi}_{1-x}\text{Sb}_x)_2\text{Te}_3$ three-dimensional topological
  insulators},}\ }\href {https://doi.org/10.1103/PhysRevB.93.125434} {\bibfield
   {journal} {\bibinfo  {journal} {Phys. Rev. B}\ }\textbf {\bibinfo {volume}
  {93}},\ \bibinfo {pages} {125434} (\bibinfo {year}
  {2016}{\natexlab{b}})}\BibitemShut {NoStop}%
\bibitem [{\citenamefont {{Schumann}}\ \emph {et~al.}(2018)\citenamefont
  {{Schumann}}, \citenamefont {{Meyer}}, \citenamefont {{Mussler}},
  \citenamefont {{Kampmeier}}, \citenamefont {{Gr{\"u}tzmacher}}, \citenamefont
  {{Schmoranzerova}}, \citenamefont {{Braun}}, \citenamefont {{Kampfrath}},
  \citenamefont {{Walowski}},\ and\ \citenamefont
  {{M{\"u}nzenberg}}}]{Thomas2018}%
  \BibitemOpen
  \bibfield  {author} {\bibinfo {author} {\bibfnamefont {T.}~\bibnamefont
  {{Schumann}}}, \bibinfo {author} {\bibfnamefont {N.}~\bibnamefont {{Meyer}}},
  \bibinfo {author} {\bibfnamefont {G.}~\bibnamefont {{Mussler}}}, \bibinfo
  {author} {\bibfnamefont {J.}~\bibnamefont {{Kampmeier}}}, \bibinfo {author}
  {\bibfnamefont {D.}~\bibnamefont {{Gr{\"u}tzmacher}}}, \bibinfo {author}
  {\bibfnamefont {E.}~\bibnamefont {{Schmoranzerova}}}, \bibinfo {author}
  {\bibfnamefont {L.}~\bibnamefont {{Braun}}}, \bibinfo {author} {\bibfnamefont
  {T.}~\bibnamefont {{Kampfrath}}}, \bibinfo {author} {\bibfnamefont
  {J.}~\bibnamefont {{Walowski}}},\ and\ \bibinfo {author} {\bibfnamefont
  {M.}~\bibnamefont {{M{\"u}nzenberg}}},\ }\bibfield  {title} {\enquote
  {\bibinfo {title} {{Observation of spin Nernst photocurrents in topological
  insulators}},}\ }\href@noop {} {\ ,\ \bibinfo {pages} {arXiv:1810.12799}
  (\bibinfo {year} {2018})}\BibitemShut {NoStop}%
\end{thebibliography}%


\providecommand{\noopsort}[1]{}\providecommand{\singleletter}[1]{#1}%
\begin{thebibliography}{1}%
\makeatletter
\providecommand \@ifxundefined [1]{%
 \@ifx{#1\undefined}
}%
\providecommand \@ifnum [1]{%
 \ifnum #1\expandafter \@firstoftwo
 \else \expandafter \@secondoftwo
 \fi
}%
\providecommand \@ifx [1]{%
 \ifx #1\expandafter \@firstoftwo
 \else \expandafter \@secondoftwo
 \fi
}%
\providecommand \natexlab [1]{#1}%
\providecommand \enquote  [1]{``#1''}%
\providecommand \bibnamefont  [1]{#1}%
\providecommand \bibfnamefont [1]{#1}%
\providecommand \citenamefont [1]{#1}%
\providecommand \href@noop [0]{\@secondoftwo}%
\providecommand \href [0]{\begingroup \@sanitize@url \@href}%
\providecommand \@href[1]{\@@startlink{#1}\@@href}%
\providecommand \@@href[1]{\endgroup#1\@@endlink}%
\providecommand \@sanitize@url [0]{\catcode `\\12\catcode `\$12\catcode
  `\&12\catcode `\#12\catcode `\^12\catcode `\_12\catcode `\%12\relax}%
\providecommand \@@startlink[1]{}%
\providecommand \@@endlink[0]{}%
\providecommand \url  [0]{\begingroup\@sanitize@url \@url }%
\providecommand \@url [1]{\endgroup\@href {#1}{\urlprefix }}%
\providecommand \urlprefix  [0]{URL }%
\providecommand \Eprint [0]{\href }%
\providecommand \doibase [0]{https://doi.org/}%
\providecommand \selectlanguage [0]{\@gobble}%
\providecommand \bibinfo  [0]{\@secondoftwo}%
\providecommand \bibfield  [0]{\@secondoftwo}%
\providecommand \translation [1]{[#1]}%
\providecommand \BibitemOpen [0]{}%
\providecommand \bibitemStop [0]{}%
\providecommand \bibitemNoStop [0]{.\EOS\space}%
\providecommand \EOS [0]{\spacefactor3000\relax}%
\providecommand \BibitemShut  [1]{\csname bibitem#1\endcsname}%
\let\auto@bib@innerbib\@empty
\bibitem [{\citenamefont {Shin}\ \emph {et~al.}(2016)\citenamefont {Shin},
  \citenamefont {Hamdou}, \citenamefont {Reith}, \citenamefont {Osterhage},
  \citenamefont {Gooth}, \citenamefont {Damm}, \citenamefont {Rellinghaus},
  \citenamefont {Pippel},\ and\ \citenamefont {Nielsch}}]{Shin2016}%
  \BibitemOpen
  \bibfield  {author} {\bibinfo {author} {\bibfnamefont {H.~S.}\ \bibnamefont
  {Shin}}, \bibinfo {author} {\bibfnamefont {B.}~\bibnamefont {Hamdou}},
  \bibinfo {author} {\bibfnamefont {H.}~\bibnamefont {Reith}}, \bibinfo
  {author} {\bibfnamefont {H.}~\bibnamefont {Osterhage}}, \bibinfo {author}
  {\bibfnamefont {J.}~\bibnamefont {Gooth}}, \bibinfo {author} {\bibfnamefont
  {C.}~\bibnamefont {Damm}}, \bibinfo {author} {\bibfnamefont {B.}~\bibnamefont
  {Rellinghaus}}, \bibinfo {author} {\bibfnamefont {E.}~\bibnamefont
  {Pippel}},\ and\ \bibinfo {author} {\bibfnamefont {K.}~\bibnamefont
  {Nielsch}},\ }\bibfield  {title} {\enquote {\bibinfo {title} {{T}he
  surface-to-volume ratio: a key parameter in the thermoelectric transport of
  topological insulator $\text{Bi}_2\text{Se}_3$ nanowires},}\ }\href
  {https://doi.org/10.1039/C6NR01716A} {\bibfield  {journal} {\bibinfo
  {journal} {Nanoscale}\ }\textbf {\bibinfo {volume} {8}},\ \bibinfo {pages}
  {13552--13557} (\bibinfo {year} {2016})}\BibitemShut {NoStop}%
\end{thebibliography}%

\end{document}